\documentclass[aps,amssymb,10pt,showpacs,showkeys,letterpaper]{revtex4-1}
\usepackage{graphicx}
\usepackage{amsmath}
\usepackage{epsfig}
\usepackage{bm}

\newcommand{\be}{\begin{equation}}
\newcommand{\ee}{\end{equation}}
\newcommand{\beq}{\begin{eqnarray}}
\newcommand{\eeq}{\end{eqnarray}}

\begin{document}

\title{Null geodesics in the Reissner-Nordstr\"{o}m Anti-de Sitter black holes}

\author{ Norman Cruz }
 \email{norman.cruz@usach.cl}
\affiliation{\it Departamento de F\'{\i}sica, Facultad de Ciencia,
Universidad de Santiago de Chile,\\ Casilla 307, Santiago 2, Chile}

\author{ Marco Olivares }
 \email{marco.olivares@ucv.cl}
\author{ Joel Saavedra }
 \email{joel.saavedra@ucv.cl}

\affiliation{\it Instituto de F\'{\i}sica, Pontificia Universidad de
Cat\'{o}lica de Valpara\'{\i}so,\\ Av. Universidad 330, Curauma,
Valpara\'{\i}so , Chile}

\author{ J. R. Villanueva }
 \email{jrvillanueval@uta.cl}
\affiliation{\it Departamento de F\'{\i}sica, Facultad de Ciencias, Universidad de Tarapac\'{a},\\
 Av. General Vel\'{a}squez 1775, Arica, Chile}

\date{\today}

\begin{abstract}
In this work we address the study of null geodesics in the
background of Reissner-Nordstr\"{o}m Anti de Sitter black holes.
We compute the exact trajectories in terms of elliptic functions
of Weierstrass, obtaining a detailed description of the orbits in
terms of charge, mass and the cosmological constant. The
trajectories of the photon are classified using the impact
parameter.
\end{abstract}



\maketitle

\section{Introduction}

From the first time that cosmological constant appears in the
literature \cite{einstein1} has been played  different and
important  roles in gravitational physics. In the last time the
cosmological constant is one of the most probable candidate to
explain the late acceleration of the Universe and solve one of the
mystery of the current observation, the called Dark Energy problem
\cite{Riess:2004nr, Peebles:2002gy, Copeland:2006wr}.
Additionally, spacetime geometry, galaxy peculiar velocities,
structure formation, and early Universe physics, supports, in many
of these cases, a flat Universe model with the presence of a
cosmological constant~\cite{Pe2}.

The above evidences have motivated to introduced spherical
symmetric spacetimes with cosmological constant in order to study
his effects as vacuum energy such it is predicted by Einstein's
gravity. The study of freely moving particles and photons in the
four-dimensional spherically symmetric space-times as attracted a
wide attention due that these spaces can described the geometry of
a family of black holes, stars and planets, which can appears in
relevant astrophysical scenarios. In the static uncharged case,
described by the Schwarzschild metric, the above study is the key
to understand various important physical phenomena: planetary
motions, gravitational lensing, radar delay, etc. The inclusion of
charge leads to the Reissner-Nordstr\"{o}m (RN) space-time. The
astrophysical importance of this solution has been perhaps not
enough considered in the literature. Some already classical
investigations has pointed out that because of a much more
frequent escape of electrons, the stars would achieve a positive
electric charge, leading to a global electrostatic field
~\cite{Shvartsman},~\cite{Punsly},~\cite{Neslusan}. For rotating
black holes endowed with electromagnetic structure, Damour and
Ruffini~\cite{Damour} pointed out the existence of the vacuum
polarization process in order to explain the Gamma Ray Bursts. See
also~\cite{others}.

The studies of moving particles and photons in this kind of
spacetime have led to determine the geodesic structure of Kottler
spacetimes~\cite{kottler}. In~\cite{jak}, timelike geodesics for
positive cosmological constant were investigated. Using the
effective potential radial null geodesic were studied in the
background of Reissner-Nordstrom-de Sitter and Kerr-de Sitter
spacetime~\cite{Stuchlik}. The geodesic structure of Schwarzschild
Anti de Sitter  (SAdS) spacetime can be found in~\cite{K-W,
Kraniotis, Hledik2, COV}. Neutral particles motion in a RN black
hole with non-zero cosmological constant was studied
in~\cite{Hledik}. Furthermore, the analytical solutions of the
geodesic equation of massive test particles in higher dimensional
SAdS, RN, and Reissner Nordstr\"{o}m (anti) de Sitter (RNAdS),
spacetimes were found in~\cite{Hackmann:2008tu, Hackmann:2008zz,
Hackmann:2008zza}, given the complete solutions and a
classification of the possible orbits in these geometries in term
of Weierstrass functions. Also, the equatorial circular motion in
Kerr- de Sitter spacetime is studied in~\cite{Slany,
Pugliese:2011xn}. Motion of neutral test particles along circular
orbits in the R-N spacetime were investigated in
~\cite{Pugliese:2010he}. Recently, in~\cite{Pugliese:2011py,
Olivares:2011xb}, were analyzed the motions of charged test
particles in RN and RNAdS space time. In this article we are
interested in study of null geodesics in the background of a RNAdS
black hole.

The paper is outlined as follows. In section II, we present the
analytical solutions for null geodesic in terms of elliptic
functions of  Weierstrass. We also discuss the effective potential
and the kind of orbits allowed in this spacetime, including the
simple case of radial motion. The gravitational bending of the
light is evaluated in terms of the RNAdS black hole parameters.
Finally, in section III, we summarize our results.

\section{Photons in the Reissner-Nordstr\"{o}m Anti-de Sitter spacetime}

\begin{figure}[!h]
 \begin{center}
   \includegraphics[width=115mm]{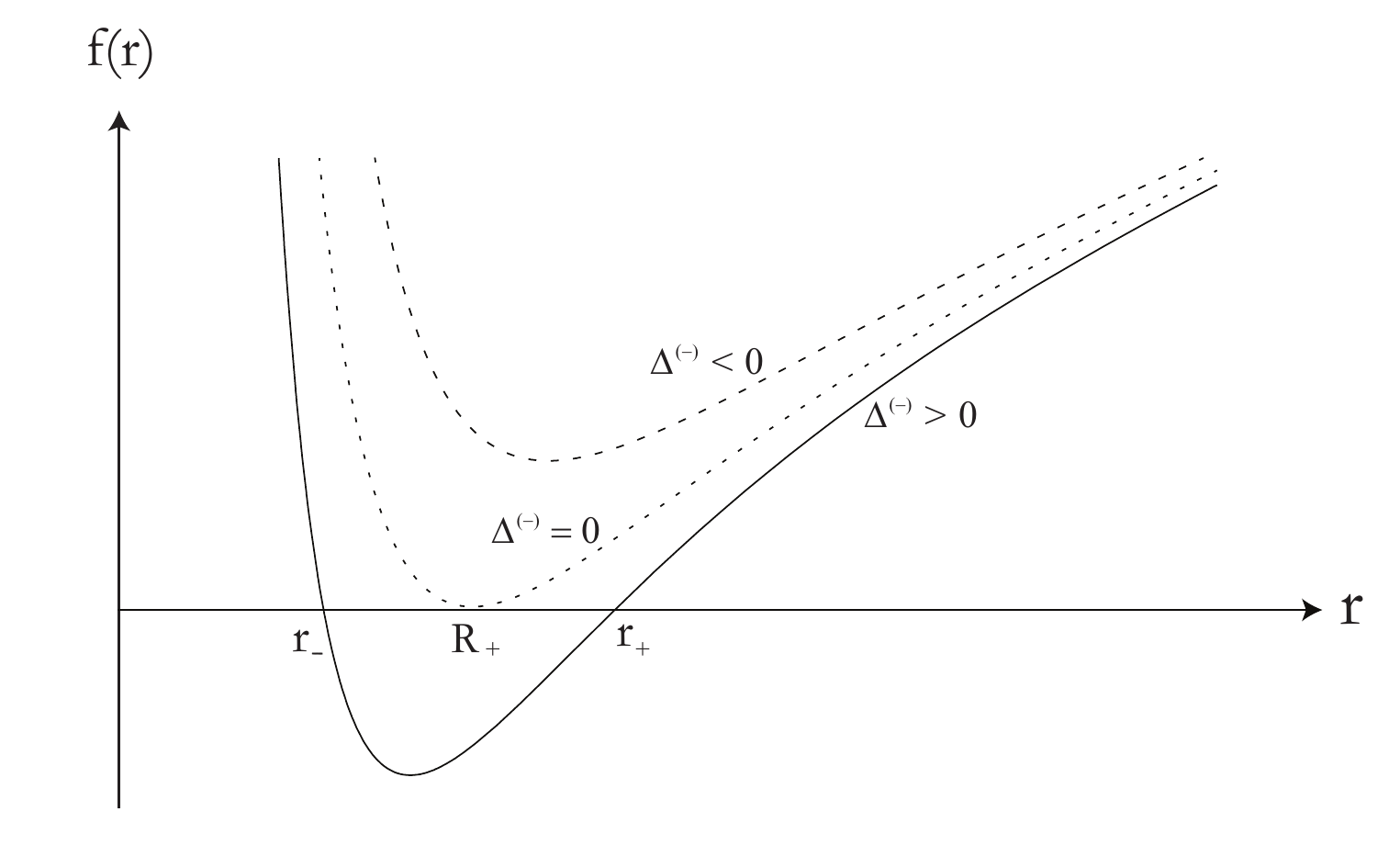}
 \end{center}
 \caption{ The lapse function for a RNAdS spacetime.
  Where the upper curve corresponds to a naked singularity, intermediate curve corresponds to an extremal black hole and the lower curve corresponds to a black hole with two horizons.}
 \label{fig1}
\end{figure}

First at all, we present a brief description of the basic aspects
of the exterior spacetime of spherical static charged black hole
embedding in a background of negative cosmological constant
$\Lambda\equiv -3/\ell^2$ (RNAdS black hole) whose metric in
Schwarzschild coordinates ($t, r, \theta, \phi$) is
\begin{equation}
ds^{2}=-f(r)dt^{2}+\frac{dr^{2}}{f(r)}+r^{2}(d\theta^{2}+\sin^{2}\theta
d\phi^{2}), \label{ra.1}
\end{equation}
where coordinates are defined in the following range: $-\infty \leq t \leq \infty$,
$r\geq0$, $0\leq\theta\leq\pi$ and $0\leq\phi\leq 2\pi$. The corresponding lapsus function $f(r)$ is
\begin{equation}
f(r)=1-\frac{2M}{r}+\frac{Q^{2}}{r^{2}}+
\frac{ r^{2}}{\ell ^{2}}\equiv \frac{F(r)}{\ell^2 r^2},
\label{ra.2}
\end{equation}
where $M$ and $Q$ represents the black hole mass and the black hole charge, respectively.
The characteristic polynomial  of RNAdS spacetime can be written as
\begin{equation}
F(r)=r^{4}+\ell^{2}(r^{2}-2Mr+Q^{2})=(r-r_+)(r-r_-)(r-r_1)(r-r_2),
\label{cp11}
\end{equation}
where $r_+$  and $r_-$  are the event horizon and Cauchy horizon. Finally ($r_1$, $r_2$) are a complex pair without physical meaning, thus
\begin{equation}
r_{\pm}=R_{+}\pm\ell\sqrt{\Delta^{(-)}}\qquad r_{1}=-R_{+}+i\ell\sqrt{\Delta^{(+)}}= r_{2}^{\ast},
\label{ra.3}
\end{equation}
where
\begin{equation}
R_{+}=\frac{\ell}{\sqrt{6}}\left\{\sqrt{1+\frac{12 Q^{2}}{\ell^{2}}}\cosh\left[\frac{1}{3}\cosh^{-1}
\left(\frac{4+ \frac{54M^{2}}{\ell^{2}}-3\left(1+ \frac{12 Q^{2}}{\ell^{2}}\right)}{\left(1+\frac{12 Q^{2}}{\ell^{2}}\right)^{3/2}}
\right)\right]-1 \right\}^{1/2},
\label{ra.4}
\end{equation}
and
\begin{equation}
\Delta^{(\pm)}=\frac{M}{2R_{+}}\pm\frac{R_{+}^{2}}{\ell^{2}}\pm\frac{1}{2}.
\label{ra.5}
\end{equation}
In this way we obtain a black hole with two horizons if
 $\Delta^{(-)}>0$; a naked singularity if $\Delta^{(-)}<0$, and an extremal black hole if $\Delta^{(-)}=0$ ( FIG.\ref{fig1} shows this behavior in detail for the lapse function).

The Lagrangian for photons in metric (\ref{ra.1}) is given by
\begin{equation}
2\mathcal{L}=-f(r)\dot{t}^{2}+\frac{\dot{r}^{2}}{f(r)}+r^{2}(\dot{\theta}%
^{2}+\sin^{2}\theta \dot\phi^2)\equiv 0,
\label{ra.7}
\end{equation}
where dot notes derivative respect to proper time $\tau$. The equations of motion can be obtained from
\be \dot{\Pi}_{q} - \frac{\partial \mathcal{L}}{\partial q} = 0,
\label{ra.8} \ee
where $\Pi_{q} = \partial \mathcal{L}/\partial \dot{q}$
are the conjugate momenta to coordinate $q$.
Since the Lagrangian is independent of ($t, \phi$) the
corresponding conjugate momenta are conserved, therefore
\begin{equation}
\Pi_{t} = -f(r) \dot{t} = - \sqrt{E}, \qquad
\textrm{and}\qquad \Pi_{\phi}
= r^{2}\sin^{2}\theta \dot{\phi} = L,
\label{ra.9}
\end{equation}
where $E$ and $L$ are constant of motion. From the equation of motion for $\theta$, we have
\begin{equation}
\frac{d(r^{2}\dot\theta)}{d\tau} =
r^{2}\sin\theta \cos\theta\dot\phi^{2}.
\label{ra.10}
\end{equation}
Without lack of generality we consider that the motion is developed in the invariant plane
 $\theta = \pi/2$, and our starting equation of motion refereed to coordinates $r$ and $\tau$ is
\begin{equation}
\left(\frac{dr}{d\tau}\right)^{2}= E-V(r),
\label{f.1}
\end{equation}
where $V(r)$ corresponds to the effective potential defined by
\begin{equation}
V(r)=f(r)\frac{L^{2}}{r^{2}}=\left(1-\frac{2M}{r}+\frac{Q^{2}}{r^{2}}+\frac{
r^{2}}{\ell^{2}}\right)\frac{L^{2}}{r^{2}}.
\label{f.2}
\end{equation}
In order to describe  the photon movement we shall use the  impact parameter
 $b\equiv\frac{L}{\sqrt{E}}$,
using the fact that $\dot r = (L/r^2)dr/d\phi$ in (\ref{f.1}) we obtain,
\begin{equation}
\left(\frac{dr}{d\phi}\right)^{2}=r^{4}\left(\frac{1}{b^{2}}
-\frac{1}{\ell^{2}}\right)-r^{2}+2Mr-Q^{2}.
\label{f.4}
\end{equation}
Dynamical description is completed adding the differential radial equation related to coordinate time
\begin{equation}
\left(\frac{dr}{dt}\right)^{2}=
f^{2}(r)\left(1-f(r)\frac{b^{2}}{r^{2}}\right).
\label{f.5}
\end{equation}
It is important to note that for photons the contribution of
cosmological constant to the effective potential is a
constant (\ref{f.2}). Then it does not functionally affect
description of the motion in a proper system (\ref{f.1}). However,
it has an influence in the motion through the coordinate time
(\ref{f.5}). In this way, photons motion is completely described
by the Eqs. (\ref{f.1}, \ref{f.4}, \ref{f.5}) and Eq.
(\ref{ra.9}). Following sections are devoted to a detailed
study of this motions.


\subsection{Radial Motion}
\begin{figure}[!h]
 \begin{center}
   \includegraphics[width=100mm]{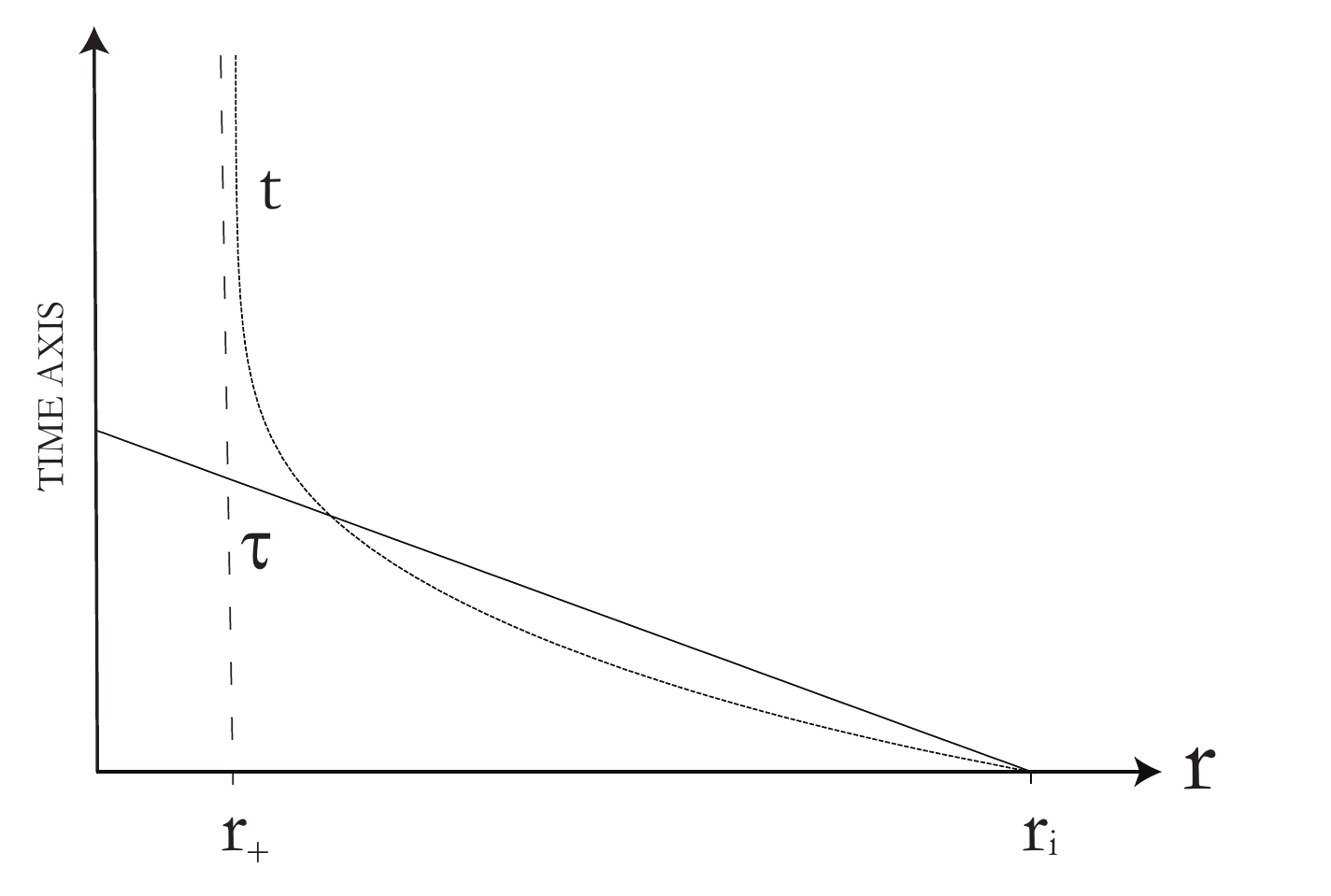}
 \end{center}
 \caption{Radial motion of the test particle for the proper time, $\tau$ (solid curve),
and the coordinate time, $t$ (dashed curve). Like the
Schwarzschild's solution, photons cross the event horizon in a
finite proper time and approaches asymptotically to this in terms
of the coordinate time.}
 \label{fig2}
\end{figure}

Radial motion corresponds to a trajectory with null angular
momentum, and we can have photons that move away of singularity
and other are doomed to fall to the singularity. From Eqs.
(\ref{f.1}) and (\ref{f.5}) we obtain
\begin{equation}
\frac{dr}{d\tau}=\pm \sqrt{E},
\label{mr.1}
\end{equation}
and
\begin{equation}
\frac{dr}{dt}=\pm\frac{F(r)}{\ell^{2}r^{2}},
\label{mr.2}
\end{equation}
 respectively, and the sign $+$ ($-$) corresponds to photons  that to go away (to get closer) from $r_+$.
If we considered that photons are in $r=r_i$
when $t=\tau=0$ and then they approach to $r_+$,
from Eq. (\ref{mr.1}) we obtain the well known result of Schwarzschild black hole
\begin{equation}
r(\tau)=r_i - \sqrt{E}\tau,
\label{mr.3}
\end{equation}
\noindent then it takes a proper time $\tau_+ = (r_i - r_+)/E$ to reach the event horizon. In order to integrated (\ref{mr.2}) we defined $F(r)=(r-r_+)(r-r_-)g(r)$, where
\begin{equation}
g(r)=(r-r_{1})(r-r_{2})
=r^{2}+2R_{+}r+[\ell^{2}+3R^{2}_{+}+(r_{+}-R_{+})^{2}].
\label{mr.4}
\end{equation}
Then we obtain
\begin{equation}
t(r)= \ln\left[\left(\frac{r-r_{+}}{r_{i}-r_{+}}\right)^{\alpha_1}
\left(\frac{r-r_{-}}{r_{i}-r_{-}}\right)^{\alpha_2}
\left(\frac{g(r)}{g(r_{i})}\right)^{\alpha_3}\right]+
\beta_1 \left[\arctan\left(\frac{r+R_{+}}{\sqrt{\ell^{2}+3R^{2}_{+}-r_{+}r_{-}}}\right)-
\arctan\left(\frac{r_i+R_{+}}{\sqrt{\ell^{2}+3R^{2}_{+}-r_{+}r_{-}}}\right)\right],
\label{mr.5}
\end{equation}
where the constants are given by
$$
\alpha_1=\frac{\ell^{2}r^{2}_{+}}{(r_{+}-r_{-})g(r_{+})},\quad
\alpha_2=-\frac{\ell^{2}r^{2}_{-}}{(r_{+}-r_{-})g(r_{-})},\quad
\alpha_3=-\frac{\ell^{2}+4R^{2}_{+}}{g(r_{+})g(r_{-})}\ell^{2}R_{+},
$$
and
$$
\beta_1=\frac{[\ell^{2}+3R^{2}_{+}+(r_{+}-R_{+})^{2}][\ell^{2}+6R^{2}_{+}-2r_{+}r_{-}]+2R^{2}_{+}r_{+}r_{-}}
 {g(r_{+})g(r_{-})[\ell^{2}+3R^{2}_{+}-r_{+}r_{-}]^{1/2}}\ell^{2}.
$$
These solutions represents, as in the Schwarzschild case,  that photons,
in the proper time framework, can cross the event horizon in a finite time and takes and infinity
coordinate time. (see Fig. \ref{fig2})

\subsection{ Angular Motion of photons}
First, we present  brief qualitative description of the allowed angular motions for photons en the RNAdS spacetime.
\begin{itemize}
  \item \emph{Capture Zone}: If $0<b\equiv b_f <b_{c}$, photons arrive from infinite and then fall to event horizon.
  \item \emph{Critical Trajectories}. If $b=b_{c}$, photons can be stayed in one of the unstable inner circular orbit of radius  $r_{c}$. Orbit radius is independent  of $L$; value of angular momentum $L$ only affects the energy $E_{c}$. Then photons that arrive from infinity can asymptotically fall to a circle of radius $r_{c}$.
  \item \emph{Deflection Zone}. If $b_{c} <b=b_D <b_{h} $, photons fall from infinite to a minimal distance  $r=r_{D}$  and then they can back to infinite. This photons are deflected. The other allowed
trajectories correspond to photons moving at the
other side of potential barrier, this photons are doomed to fall into
the event horizon.

  \item \emph{Pascal Lima\c{c}on}. For $b =b_{h}$ trajectory of photons are represented by the Pascal Lima\c{c}on. This kind of trajectory does not appears in the Schwarzschild or
Reissner-Nordstr$\ddot{o}$m cases, then are characteristics trajectories of RNAdS spacetime.
  \item \emph{ Confinement Zone}. If $b_{h} <b= b_z <\infty$, photons fall into the event horizon from an initial distance  $r_{0}$.
\end{itemize}

\begin{figure}[!h]
 \begin{center}
   \includegraphics[width=105mm]{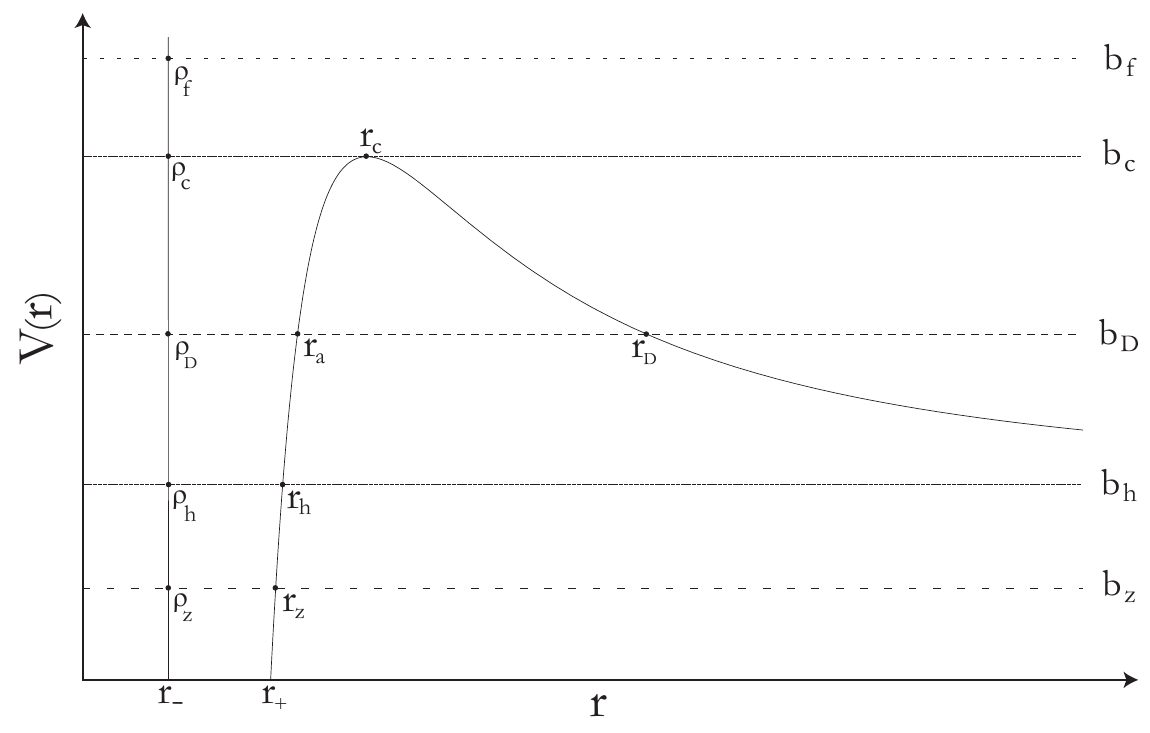}
 \end{center}
 \caption{Effective potential for null geodesics in the RNAdS spacetime. Different allowed orbits are characterized by the impact parameter $b$.}
 \label{fig3}
\end{figure}

In order to do a quantitative analysis, we define the  {\it anomalous impact parameter}
\begin{equation}
\frac{1}{b^2}-\frac{1}{\ell^2}=\left\{\begin{array}{cl} \frac{1}{B^2}, \qquad \textrm{if}\quad b < b_h,\\
\newline\\
-\frac{1}{B_{z}^{2}}, \qquad \textrm{if}\quad b=b_z > b_h,
\end{array} \right.
\label{ma}
\end{equation}
then eq. (\ref{f.4}) can be rewritten as
\begin{equation}
\left(\frac{dr}{d\phi}\right)^{2}= \frac{r^{4}-B^{2}(r^{2}-2Mr+Q^{2})}{B^{2}}\equiv \frac{P(r)}{B^2}.
\label{ma.1}
\end{equation}

\subsubsection{Captured Photons }
\begin{figure}[!h]
\begin{center}
\includegraphics[width=70mm]{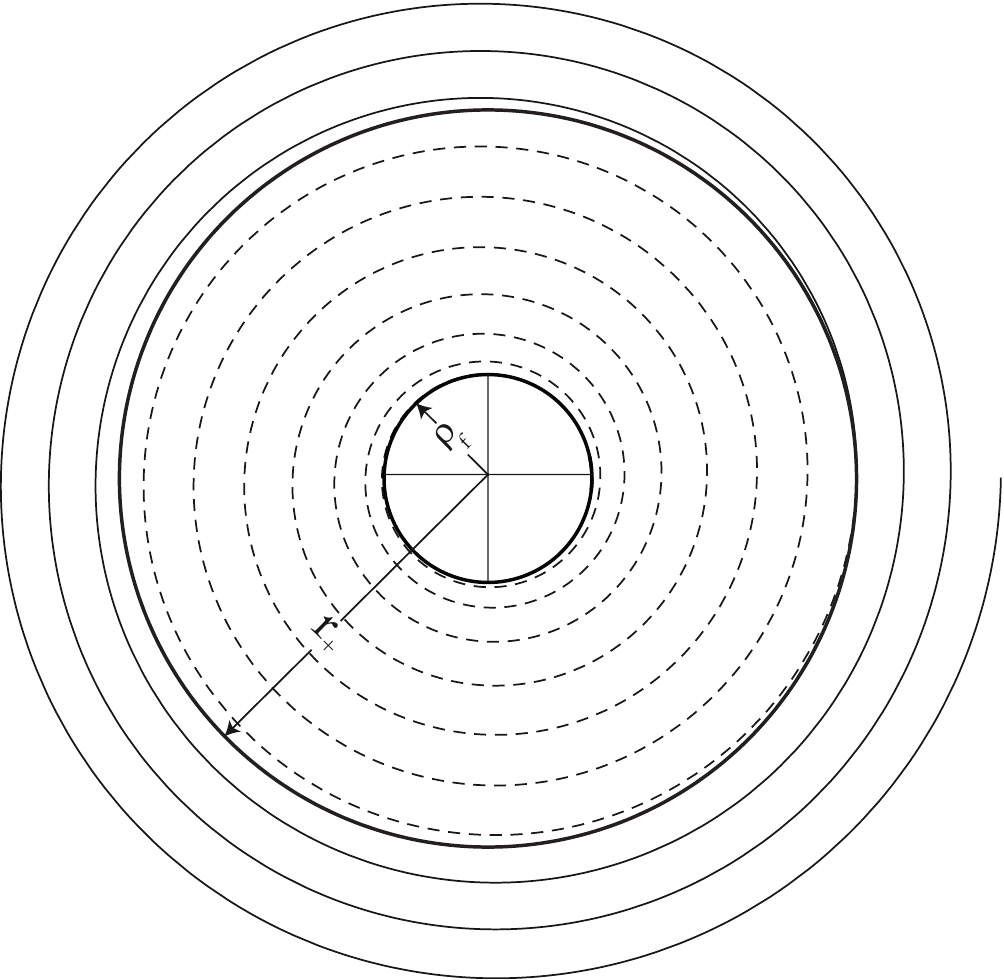}
\end{center}
\caption{Photons which cross both horizons and are reflected at
the potential barrier and emerges in another universe.}
\label{fig4}
\end{figure}

The capture zone is defined by values of impact parameter  $B=B_f$, where radial coordinate $r$ is restricted to values
$\rho_f \leq r <\infty$ (see Fig. \ref{fig3}). Therefore, characteristic polynomial has two reals roots, $\rho_f$ and $\sigma_f$ $(\rho_f >0>\sigma_f)$, and two complex roots, $r_f$ and $R_f$ ($R_f=r_{f}^{*}$), and is defined by
\begin{equation}
P_{f}(r)=(r-\rho_f)(r-\sigma_f)(r-r_{f})(r-r_{f}^{*}).
\label{fc.0}
\end{equation}
Using the change of variables $u=r-\rho_f$, and writing the
integration in the growing direction of $r$,
\begin{equation}
\phi(u)= B_f \int^{u}_{0}\frac{du}{\sqrt{u Q_{f}(u)}},
\label{fc.1}
\end{equation}
where the third degree polynomial  $Q_f(u)$ is given by
\begin{equation}
Q_{f}(u)=(u+u_{1})(u+u_{2})(u+u_{2}^*),
\label{fc.2}
\end{equation}
\noindent and his constants are given by  $u_{1}=\rho_f-\sigma_f$ y $u_{2}=\rho_f - r_f$.
Now using the following constants
$\alpha_f= (u_{1})^{-1}+(u_{2})^{-1}+(u_{2}^*)^{-1}$,
$\beta_f=(u_{1}u_{2})^{-1}+(u_{1}u_{2}^*)^{-1}+|u_{2}|^{-2}$, $\gamma_f=1/C_f^2$,
$\kappa_f=\frac{C_f}{B_f}$ y $C_f=|u_{2}|\sqrt{u_{1}}$,
and doing the change of variable $u^{-1} = 4 U - \alpha_f/3$, we obtain the quadrature form
\begin{equation}
\kappa_f \phi= \int^{\infty}_{U}\frac{dU}{\sqrt{4U^{3}-g_{2_f} U-g_{3_f}}},
\label{fc.3}
\end{equation}
where the invariants are given by,
\begin{equation}
g_{2_f}=\frac{1}{4}\left(\frac{\alpha_f^{2}}{3}-\beta_f\right),\quad \textrm{y}\quad
g_{3_f}=\frac{1}{16}\left(-\gamma_f-\frac{2\alpha_f^{3}}{27}+\frac{\alpha_f\beta_f}{3}\right).
\label{fc.4}
\end{equation}
We found the solution of (\ref{fc.3}) in terms of the elliptic
function of  Weierstrass, $U=\wp(\kappa_f \phi)$, and we can
perform the inversion and found the relation between radial and
polar coordinates $r$ and $\phi$ (see Fig. \ref{fig4}),
\begin{equation}
r(\phi)=\rho_f+\frac{1}{4\wp(\kappa_f \phi)-\frac{\alpha_f}{3}}.
\label{fc.6}
\end{equation}
This equation represents the orbits of captured photons.
\subsubsection{Critical Orbits of Photons}
\begin{figure}[!h]
 \begin{center}
   \includegraphics[width=75mm]{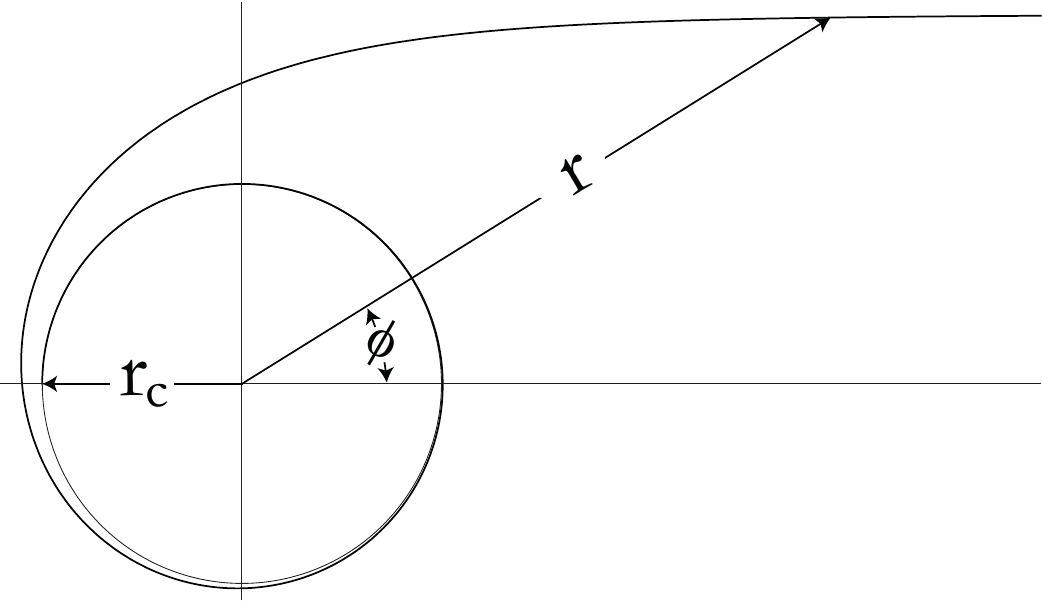}
 \end{center}
 \caption{Critical first kind orbit. Photons are coming from the infinity and approach asymptotically to the unstable circular orbit.}
 \label{fig5}
\end{figure}

This critical orbit is a consequence of the potential has a maximum in $r_c$ (see Fig: \ref{fig3}),
and the photons have a impact parameter $B=B_c$. This critical distance can be obtained by the equation
$\left( \frac{dV(r)}{dr}\right)_{r=r_c}=0$, and yields
\begin{equation}
r_{c}=\frac{3M}{2} \left[ 1+\sqrt{1-\frac{8Q^{2}}{9M^{2}}}\right].
\label{mc.2}
\end{equation}
The unstable circular orbit for photons is independent of the
cosmological constant and only depends of the mass and charge of
the black hole. In this form, the characteristic polynomial has
his general form,
\begin{equation}
P_{c}(r)=(r-\rho_{c})(r-\sigma_{c})|r-r_{c}|^{2}.
\label{mc.04}
\end{equation}
We can also obtain the value of  $B_c$
\begin{equation}
\frac{1}{B_c^{2}} =\frac{1}{r_{c}^{2}}
\left(1-\frac{2M}{r_{c}}+\frac{Q^{2}}{r_{c}^{2}}\right).
\label{mc.3}
\end{equation}
\paragraph{Critical Motion of First Kind}
The first kind of critical trajectories corresponds to the motion of photons
that came from infinite and reach asymptotically  to unstable circular orbit,$r_c < r$.
In this way, the characteristic polynomial for first kind trajectories, (\ref{mc.04}),
is given by
\begin{equation}
P_{c}(r)=(r-\rho_{c})(r-\sigma_{c})(r-r_{c})^{2}.
\label{mc.4}
\end{equation}
Using the change of variables $y=r-r_{c}$ in Eq. (\ref{ma.1}) we
obtain
\begin{equation}
\phi(y)= B_c \int^{y}_{\infty}\frac{-dy}{y\sqrt{ Q_{c}(y)}},
\label{mc.5}
\end{equation}
where second degree polynomial is given by
$Q_{c}(y)=(y+y_{1})(y+y_{2})$, whose constant are defined as $y_{1}=r_{c}-\rho_{c}$ and $y_{2}=r_{c}-\sigma_{c}$.
Finally using the change $Y=1/y$,
and defining  $\alpha_c = (y_{1})^{-1} +(y_{2})^{-1}$,
$\beta_c = (y_{1}y_{2})^{-1}$, $\kappa_c=\frac{C_c}{B_c}$ and
$C_c=\sqrt{y_{1}y_{2}}$, we obtain
\begin{equation}
\kappa_c \phi(Y)= \int^{Y}_{0}\frac{dY}{\sqrt{Y^{2}+\alpha_c Y+\beta_c}},
\label{mc.6}
\end{equation}
and
\begin{equation}
\kappa_c \phi= \ln\left(a_{_1}\sqrt{Y^{2}+\alpha_c Y+\beta_c}+a_{_1} Y + a_{_2}\right),
\label{mc.7}\end{equation}
where $a_{_1}=2(2\sqrt{\beta_c}+\alpha_c)^{-1}$ and
$a_{_2}=\alpha_c(2\sqrt{\beta_c}+\alpha_c)^{-1}$. Finally the inversion of  (\ref{mc.7}) allows to find the polar form of the motion
\begin{equation}
r(\phi)=r_{c}+\frac{A_{1}+A_{2}(e^{\kappa_c \phi}-a_{_2})}{A_3(e^{\kappa_c \phi}-a_{_2})^{2}-1},
\label{mc.8}
\end{equation}
where the constants are given by
$A_{1}=\frac{\alpha_c}{\beta_c}$, $A_{2}=\frac{2}{a_{_1}\beta_c}$ and
$A_{3}=\frac{1}{a_{_1}^{2}\beta_c}$. In Fig. \ref{fig5} we show the form of this orbit.

\begin{figure}[!h]
 \begin{center}
   \includegraphics[width=70mm]{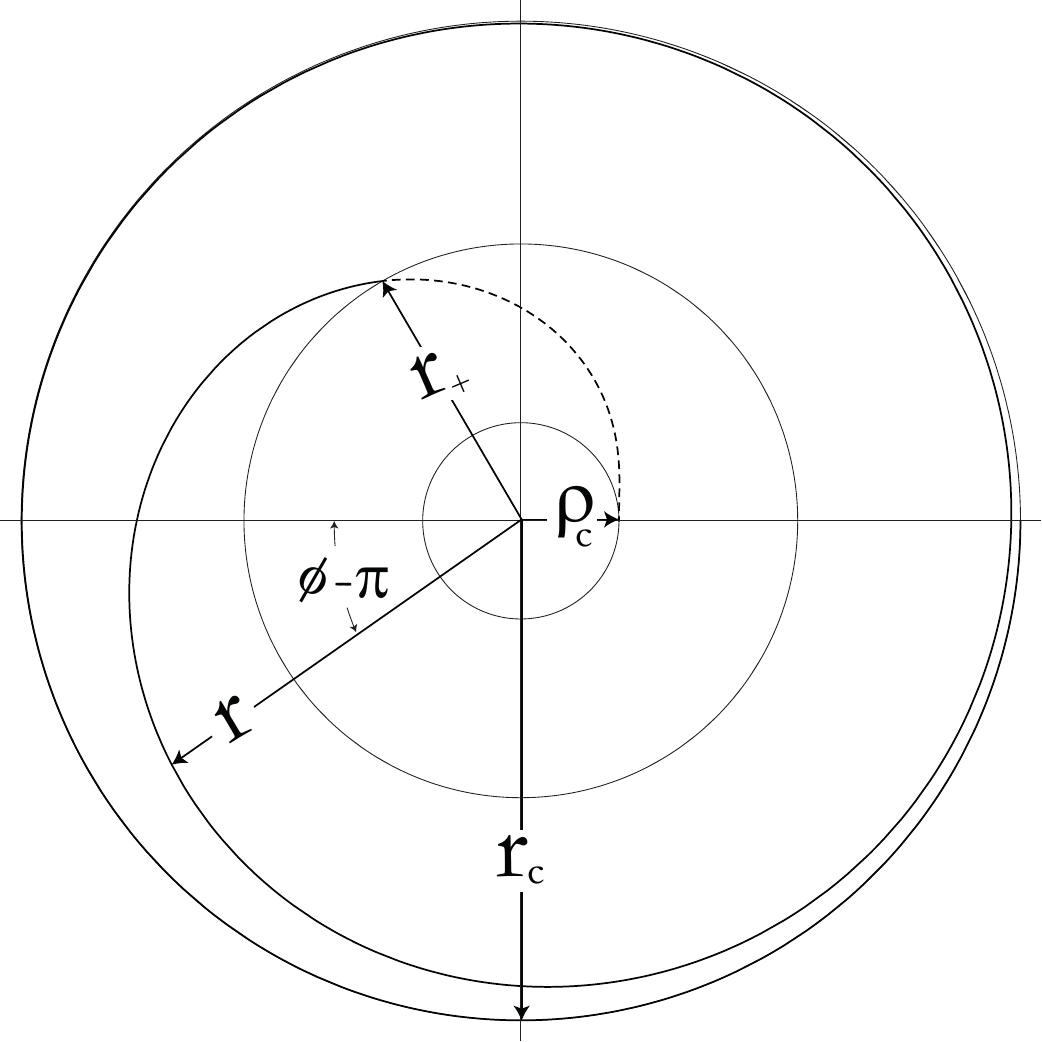}
 \end{center}
 \caption{Critical trajectory of second kind. Photons are coming from distances
 lower than the critical radius and approach asymptotically to the critical radius.}
 \label{fig6}
\end{figure}

\paragraph{Critical motion of second kind}
This motion corresponds to photons going to the unstable circular orbit
from distance greater than the event horizon but less than critical radius
$r_{c}>r>r_+$. In this way, the characteristic polynomial given by eq. (\ref{mc.04}),
takes the form
\begin{equation}
P_{c}(r)=(r-\rho_{c})(r-\sigma_{c})(r_{c}-r)^{2}.
\label{mc.9}
\end{equation}
 Using the change of variables
$\widetilde{y}=r_{c}-r$, eq. (\ref{ma.1})
becomes from (\ref{mc.9}),
\begin{equation}
\phi(\widetilde{y})= B_c \int^{\widetilde{y}}_{\widetilde{y}_{1}}
\frac{-d\widetilde{y}}{\widetilde{y}\sqrt{ \widetilde{Q}_{c}(\widetilde{y})}},
\label{mc.10}
\end{equation}
where
$\widetilde{Q}_{c}(\widetilde{y})=(\widetilde{y}_{1}-\widetilde{y})(\widetilde{y}_{2}-\widetilde{y})$,
and the constants are the same constants that the first kind motion
$\widetilde{y}_{1}=r_{c}-\rho_{c}=y_1$, and $\widetilde{y}_{2}=r_{c}-\sigma_{c}=y_2$,
then $\widetilde{\alpha}_c = \alpha_c$,
$\widetilde{\beta}_c=\beta_c$, $\widetilde{\kappa}_c=\kappa_c$, $\widetilde{C}_c=C_c$,
$\widetilde{a}_{_1}=a_{_1}$, $\widetilde{a}_{_2}=a_{_2}$,
$\widetilde{A}_{_1}=A_{_1}$ and $\widetilde{A}_{_2}=A_{_2}$.
Thus we find from eq. (\ref{mc.10}), and using $\widetilde{Y}=1/\widetilde{y}$, that
\begin{equation}
\kappa_c \phi= \ln\left(a_{_1}\sqrt{Y^{2}-\alpha_c Y+\beta_c}+a_{_1} Y - a_{_2}\right),
\label{mc.11}
\end{equation}
and his polar form (see Fig. \ref{fig6}),
\begin{equation}
r(\phi)=r_{c}+\frac{A_{1}-A_{2}(e^{\kappa_c\phi}+a_{_2})}{A_{3}(e^{\kappa_c\phi}+a_{_2})^{2}-1}.
\label{mc.12}
\end{equation}

\paragraph{Circular orbits}
Once photons travel from infinite, or from another distance  $r_i$  ($r_c> r_i>r_+$), they go asymptotically
to the unstable circular orbit. Their period is practically a constant. Then we can compute this period to respect the proper time,
\begin{equation}
T_{\tau}=\frac{2\pi r_c^{2}}{L}=\frac{9\pi M^{2}}{2L} \left[ 1+\sqrt{1-\frac{8Q^{2}}{9M^{2}}}
\right]^{2},
\label{mc.13}
\end{equation}
while the period in coordinate time is
\begin{equation}
T_{t}= \frac{2\pi r_{c}^{2}}{L }\frac{ \sqrt{E_{c}}}{f(r_c)}=
T_{\tau} \frac{ \sqrt{E_{c}}}{f(r_c)}= \frac{2\pi r_c}{\sqrt{f(r_c)}}.
\label{mc.14}
\end{equation}
The above results show that both periods do not depend of the cosmological constant.

\subparagraph{Schwarzschild Limit}
Considering the circular movement of photons, the unstable radius is $3M$, then periods to circular orbits for photons reduce to the following terms $T_{\tau}=\frac{18 \pi M^{2}}{L}$ and $T_{t} = 6\pi M
\sqrt{3}$ \cite{Shutz}.


\subsubsection{Deflection Zone}
In this case, we have that anomalous impact parameter is  $B_{_D}$, and
from Fig. \ref{fig3}, there are two possibilities for photons with this impact parameter.
In the trajectory of first class the photons came from infinite and are deflected,
approaching to a minimal distance in $r=r_{_D}$ and going away to infinite   ($r_+ < r_a<r_{_D}<r$);
in the trajectory of second class the photon motion is allowed between two distance
$r_+<r<r_a<r_{_D}$. The extremal distances $r_{_D}$ and $r_a$ are obtained from the extremal condition
$\frac{dr}{d\phi}\mid_{r_{extr}}=0$. On the other side, the polynomial of fourth degree has four real roots, and it can be written in a general form
\begin{equation}
P_{_D}(r)=|r-r_{_D}|\,|r-r_{a}|\,(r-\rho_{_D})\,(r-\sigma_{_D}).
\label{zd.2}
\end{equation}

\begin{figure}[!h]
 \begin{center}
   \includegraphics[width=105mm]{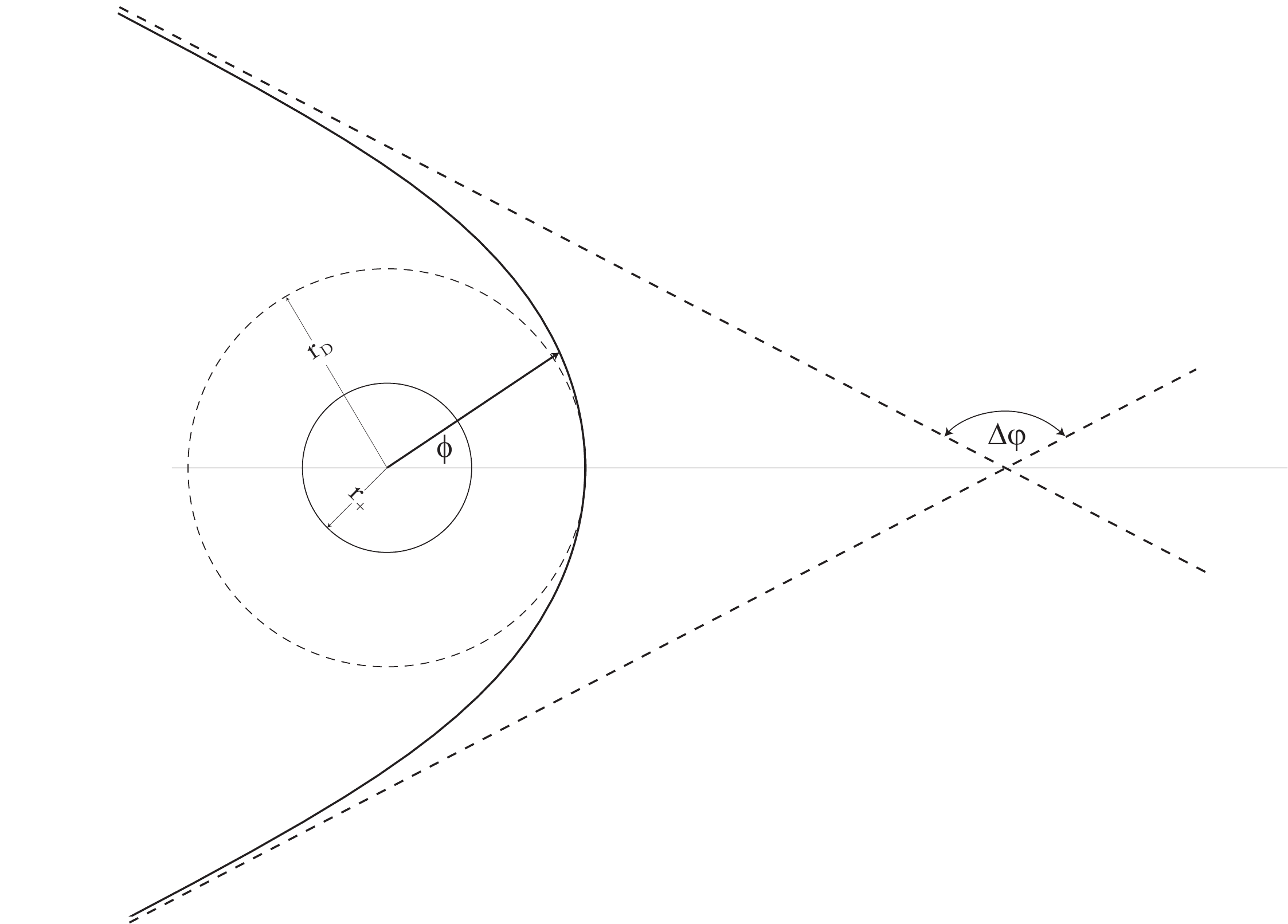}
 \end{center}
 \caption{Light deflection in the background of a RNAdS black hole. The figure shows the deflection angle $\Delta \varphi$.}
 \label{fig7}
\end{figure}

\paragraph{First class trajectory:}
This trajectory corresponds to the light deflection. We start computing the trajectory from the extremal point $r_{_D}$ and using the change of variable
$w=r-r_{D}$, we obtain
\begin{equation}
\phi(w)= B_{_D} \int^{w}_{0}\frac{dw}{\sqrt{w Q_{_D}(w)}},
\label{zd.3}\end{equation}

\noindent where third degree polynomial in the new variable is
$Q_{_D}(w)=(w_{1}+w)(w_{2}+w)(w_{3}+w)$, where
$w_{1}=r_{_D}-r_{a}$, $w_{2}=r_{_D}-\rho_{_D}$ and
$w_{3}=r_{_D}-\sigma_{_D}$. Now defining the constants
$\alpha_{_D}= (w_{1})^{-1} + (w_{2})^{-1} + (w_{3})^{-1}$,
$\beta_{_D} = (w_{1}w_{2})^{-1} + (w_{1}w_{3})^{-1}+
(w_{2}w_{3})^{-1}$, $\gamma_{_D} = C_{_D}^{-2}$,
$\kappa_{_D}=\frac{C_{_D}}{B_{_D}}$,
$C_{_D}=\sqrt{w_{1}w_{2}w_{3}}$, and doing the change of variable
$w^{-1}=4W-\alpha_{_D}/3$,  Eq. (\ref{zd.3}) give us
\begin{equation}
\kappa_{_D} \phi= \int^{\infty}_{W}\frac{dW}{\sqrt{4W^{3}-g_{2_D} W-g_{3_D}}},
\label{zd.4}
\end{equation}
where the invariants are
\begin{equation}
g_{2_D}=\frac{1}{4}\left(\frac{\alpha_{_D}^{2}}{3}-\beta_{_D}\right),\qquad \textrm{y}\qquad
g_{3_D}=\frac{1}{16}\left(\frac{\alpha_{_D}\beta_{_D}}{3}-\gamma_{_D}-\frac{2\alpha_{_D}^{3}}{27}\right),
\label{zd.5}
\end{equation}
whose solution is $W=\wp(\kappa_{_D} \phi)$. This gives the parametric relation (see Fig. \ref{fig7})
\begin{equation}
r(\phi)=r_{_D}+\frac{1}{4\wp(\kappa_{_D} \phi)-\frac{\alpha_{_D}}{3}},
\label{zd.6}
\end{equation}

The angle of deflection for photons is defined by
\begin{equation}
\Delta \varphi = 2|\phi(\infty)-\phi(r_{_D})|-\pi,
\label{zd.7}
\end{equation}
or in terms of the inverse Weierstrass function
\begin{equation}
\Delta \varphi = \frac{2}{\kappa_{_D}}\wp^{-1}\left(\frac{\alpha_{_D}}{12}\right)-\pi.
\label{zd.8}
\end{equation}

\begin{figure}[!h]
 \begin{center}
   \includegraphics[width=70mm]{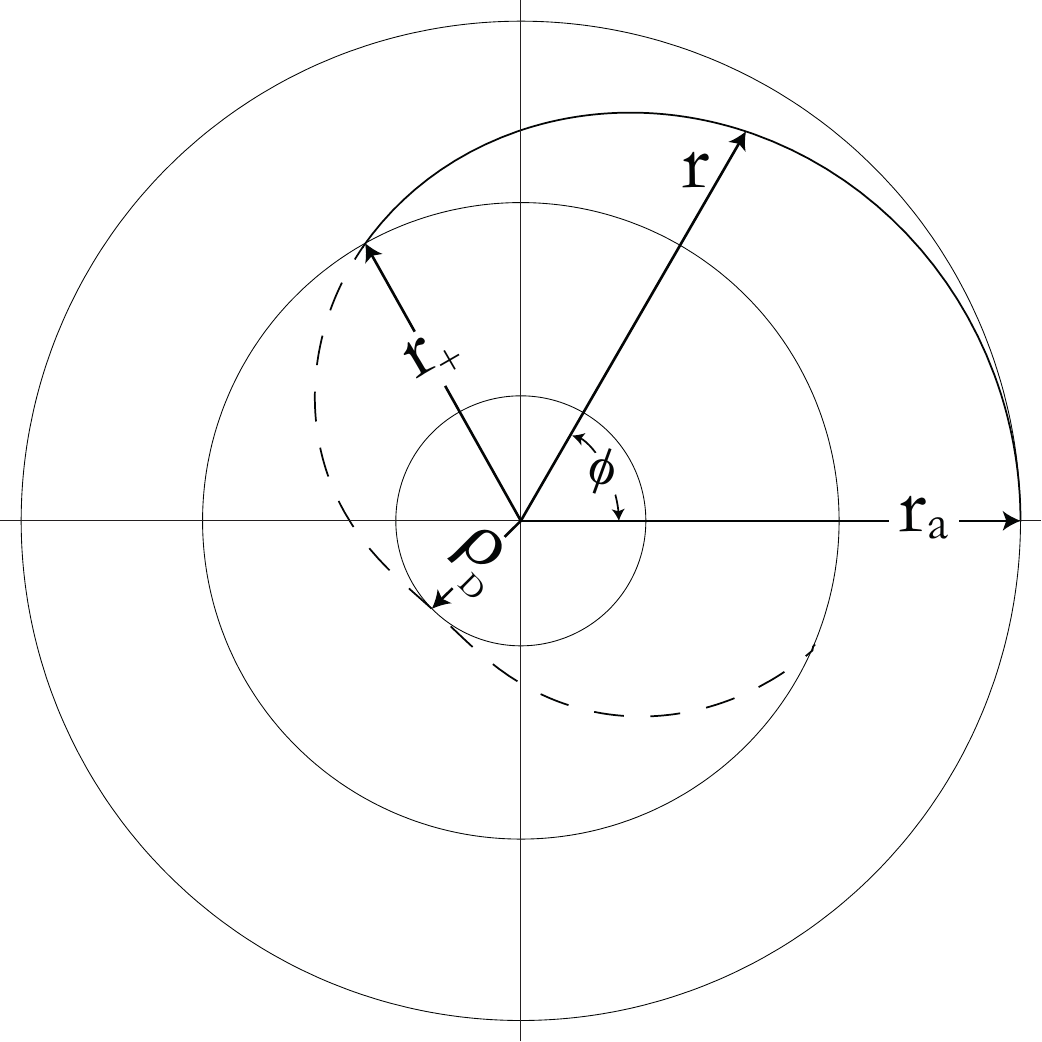}
 \end{center}
 \caption{Many-world bound orbit for null geodesics of second class.}
 \label{fig8}
\end{figure}

\paragraph{Second class trajectories:}
In this case photons have a maximal distance of remoteness, $r_a$. This distance can be chosen as the starting point of trajectory.
Using the change of variable $x=r_{a}-r$, we can obtain
\begin{equation}
\phi(x)= B_{_D} \int^{x}_{0}\frac{dx}{\sqrt{x \widetilde{Q}_{_D}(x)}},
\label{zd.9}
\end{equation}
where the cubic polynomial is given by
$\widetilde{Q}_{D}(x)=(x+x_{1})(x_{2}-x)(x_{3}-x)$, and
the corresponding constants are given by $x_{1}=r_{_D}-r_{a}$,
$x_{2}=r_{a}-\rho_{_D}$ y $x_{3}=r_{a}-\sigma_{_D}$. Beside, we use the following definition
$\widetilde{\alpha}_{_D} = -(x_{1})^{-1} + (x_{2})^{-1} + (x_{3})^{-1}$,
$\widetilde{\beta}_{_D} = (x_{1}x_{2})^{-1} + (x_{1}x_{3})^{-1} - (x_{2}x_{3})^{-1}$,
$\widetilde{\gamma}_{_D} = \widetilde{C}_{_D}^{-2}$, $\widetilde{\kappa}_{_D}=\frac{\widetilde{C}_{_D}}{B_{_D}}$
$\widetilde{C}_{_D}=\sqrt{x_{1}x_{2}x_{3}}$, and doing the following change of variable $x=4X+\alpha_D/3$
in Eq. (\ref{ma.1}), we obtain
\begin{equation}
\widetilde{\kappa}_{_D} \phi= \int^{\infty}_{X}\frac{dX}{\sqrt{4X^{3}-
\widetilde{g}_{2_D} X-\widetilde{g}_{3_D}}},
\label{zd.10}
\end{equation}
whose solution is $X=\wp(\widetilde{\kappa}_{_D} \phi)$, where the invariant are given by
\begin{equation}
\widetilde{g}_{2_D}=\frac{1}{4}\left(\frac{\alpha_{_D}^{2}}{3}+\beta_{_D}\right),\qquad \textrm{y}\qquad
\widetilde{g}_{3_D}=\frac{1}{16}\left(\frac{\alpha_{_D}\beta_{_D}}{3}-\gamma_{_D}+\frac{2\alpha_{_D}^{3}}{27}\right).
\label{zd.10}
\end{equation}
Inverting to obtain the coordinate $r$ as a function of $phi$, we obtain
the polar equation for the orbits
\begin{equation}
r(\phi)=r_{a}-\frac{1}{4\wp(\widetilde{\kappa}_{_D} \phi)+\frac{\widetilde{\alpha}_{_D}}{3}}.
\label{zd.11}
\end{equation}
The above solution is shown in Fig. \ref{fig8}.
\subsubsection{Pascal Lima\c{c}on}
\begin{figure}[!h]
 \begin{center}
   \includegraphics[width=65mm]{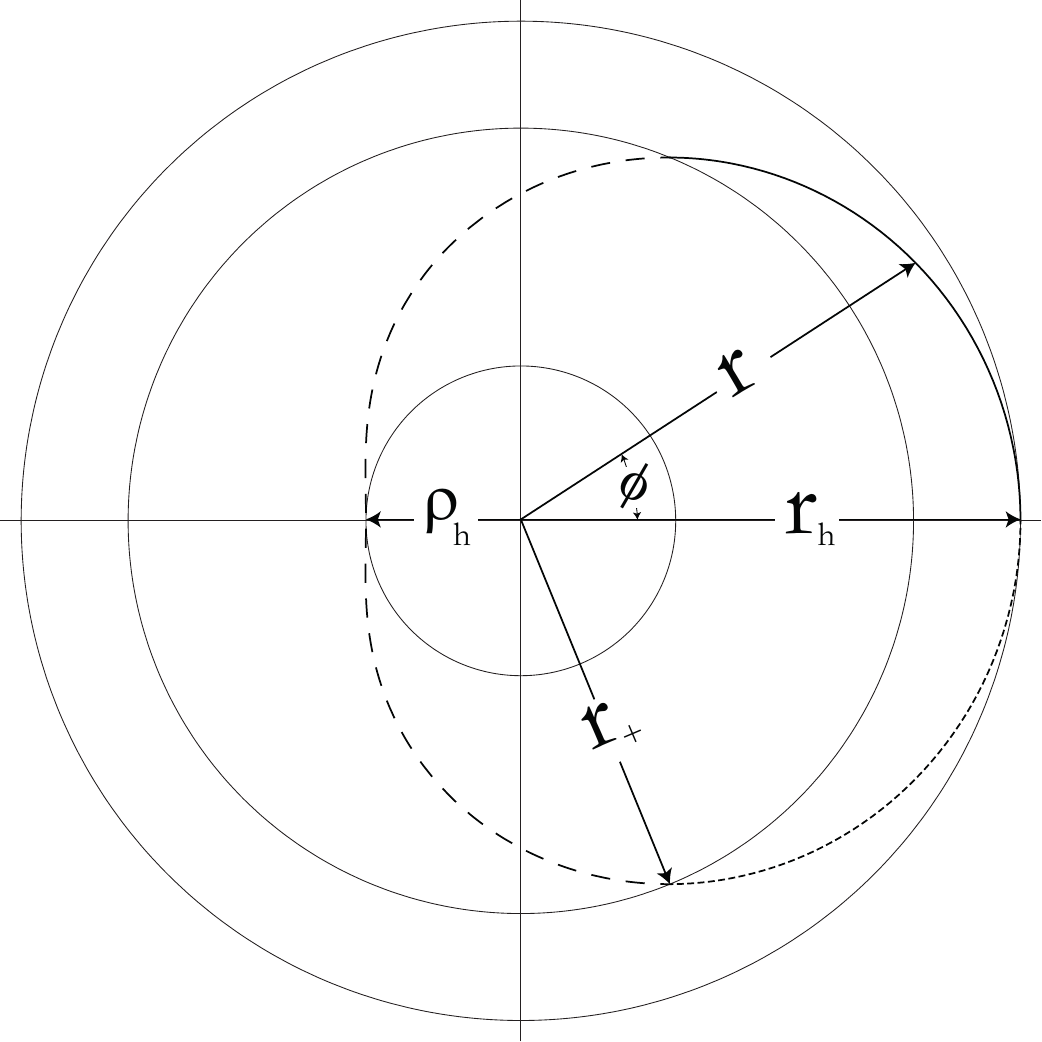}
 \end{center}
 \caption{Many-world bound orbit for null geodesics, which is described by a Pascal Lima\c{c}on. If $Q=0$, the trajectory is a cardioid.}
 \label{fig9}
\end{figure}

This kind of trajectory is an unique solution for the geodesic
problem of RNAdS. It corresponds to one epicycle curve defined by his anomalous impact parameter
 $B_h\rightarrow \infty$ ($b_h=\ell$). Therefore, radial coordinate is restricted to
$r_+<r<r_h$, and the equation of motion (\ref{ma.1})
can be written as
\begin{equation}
\phi(r)=\int_{r_h}^{r}\frac{-dr}{\sqrt{Q_h(r)}},
\label{lp.1}
\end{equation}
where the second degree polynomial is defined by
$Q_{h}(r)=(r_h-r)(r-\rho_h)$, and his roots are:
$r_{h}=M+\sqrt{M^{2}-Q^{2}}$ y $\rho_{h}=M-\sqrt{M^{2}-Q^{2}}$.
It is straightforward to find solution of Eq. (\ref{lp.1})
(see Fig. \ref{fig9}), which is given by
\begin{equation}
r(\phi)=M+\sqrt{M^{2}-Q^{2}}\cos\phi.
\label{lp.2}
\end{equation}
Note that this trajectory is an exclusive solution of black hole with
cosmological constant and it does not depend on its value

\subsubsection{Confined Photons}
\begin{figure}[!h]
 \begin{center}
   \includegraphics[width=75mm]{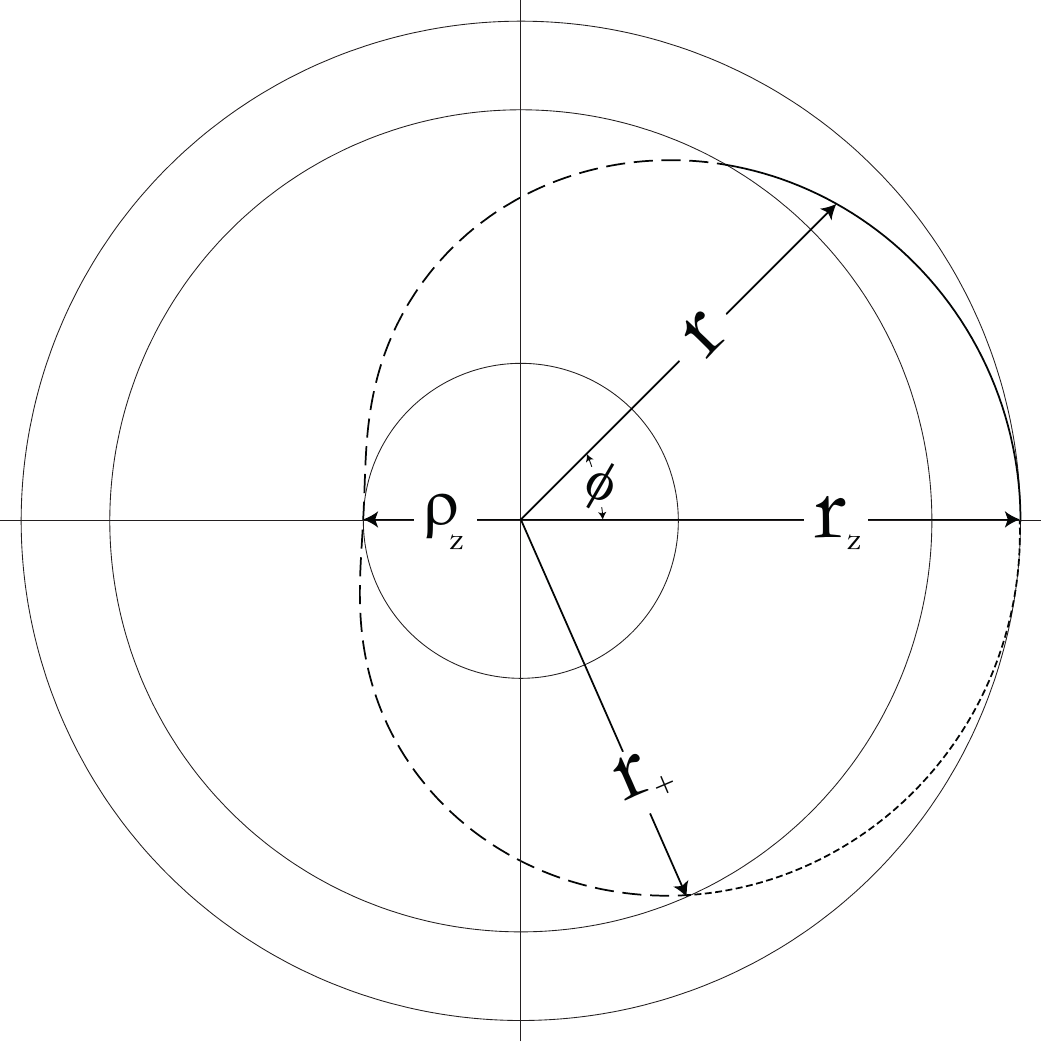}
 \end{center}
 \caption{Many-world bound orbit for confined photons}
 \label{fig10}
\end{figure}

In this case, photons have an anomalous impact parameter
$B_z$, con $b_z > b_h$. In this region radial coordinate is restricted to $r_+<r<r_z$, and the characteristic polynomial has two different real roots, $\rho_z$ y $r_z$  ($\rho_z < r_z$), beside to a conjugate pair
$r_2$ y $R_2$ ($R_2=r_2^{*}$). Using $\phi = 0$ for $r=r_z$, together which the change of variable $s=r_{z}-r$,
we obtain
\begin{equation}
\phi(r)= B_z \int^{s}_{s_{z}}\frac{ds}{ \sqrt{s Q_{z}(s)}},
\label{cp.1}
\end{equation}
where the third degree polynomial in the new variable is
$Q_z(s)=(s_{1}-s)(s_{2}-s)(s_{2}^{*}-s)$, and the constants are
$s_{1}=r_{z}-\rho_{z}$ y $s_{2}=r_{z}-r_{2}$. Using the constants
$\alpha_{z} = (s_{1})^{-1}+(s_{2})^{-1}+(s_{2}^{*})^{-1}$,
$\beta_{z} = (s_{1}s_{2})^{-1} + (s_{1}s_{2}^{*})^{-1} +
|s_{2}|^{-2}$, $\gamma_{z} = C_{z}^{-2}$,
$\kappa_{z}=\frac{C_z}{B_z}$, and $C_z=|s_2|\sqrt{s_{1}}$,
together the change of variable $s^{-1}=4S+\alpha_z/3$, we obtain
\begin{equation}
\kappa_z \phi(S)= \int^{\infty}_{S}\frac{dS}{\sqrt{4S^{3}-g_{2_z}
S-g_{3_z}}},
\label{cp.2}
\end{equation}
where the invariant are
\begin{equation}
g_{2_z}=\frac{1}{4}\left(\frac{\alpha_z^{2}}{3}-\beta_z\right),\quad
\textrm{y}\quad
g_{3_z}=\frac{1}{16}\left(\gamma_z+\frac{2\alpha_z^{3}}{27}-
\frac{\alpha_z\beta_z}{3}\right).
\label{cp.3}
\end{equation}
Again the solution of (\ref{cp.2}) is obtained in terms of the  Weierstrass function $\wp$, $S=\wp(\kappa_z \phi)$,
where the inversion give us the polar equation of confined orbit (see Fig. \ref{fig10})
\begin{equation}
r(\phi)=r_{z}-\frac{1}{4\wp(\kappa_z \phi)+\frac{\alpha_z}{3}}.
\label{cp.4}
\end{equation}
\section{Final Remarks}
We have studied the geodesic structure of RNAdS black holes analyzing the
behavior of null geodesic by means of the
effective potential which appears in the radial equation of motion. We
separated our study in two cases, radial null
geodesics and angular null geodesics. We have obtained, for the radial
case, analytical solutions for the proper and
coordinate time. Our solutions are very similar to Schwarzschild case where
in the proper time framework photons
can cross the event horizon and in a finite but is infinity in the
coordinate time framework. In the angular motions of photons we have
found that are different kind of motions
depending of the impact parameter (b) of the orbits.
We have obtained five different kinds of motions for captured
photons, which arrive from infinite and fall into the event horizon.
The trajectories of these null geodesics have
been given in terms of elliptic function of Weierstrass.
In second place, we have also found photons following critical orbits,
representing movements that came  from infinite and can asymptotically to
fall to a circle.
Third, we have found  the deflection zone that represents photons falling
from infinite to a minimal distance and then going back to infinite again.
The fourth kind of orbit is described by Pascal
Lima\c{c}on. This trajectory is an exclusive solution of black hole
which cosmological constant and has the particularity that does not depend
of the value of cosmological constant.
Finally, our last kind of orbit represents confined photons moving in the
region $r_+<r<r_z$.

\begin{acknowledgments}
J.S. was supported by COMISION NACIONAL DE CIENCIAS Y TECNOLOGIA
through FONDECYT Grant 1110076, 1090613 and 1110230. Also J.S.
work was partially supported by PUCV DI-123.713/2011. N.C.
acknowledges the support to this research by CONICYT through Grant
No. 1110840. M.O. and J. V. acknowledge the hospitality of the
Physics Department of Universidad de Santiago de Chile. M.O.
was supported by PUCV through Proyecto DI Postdoctorado 2011. J. V.
was supported by Universidad de Tarapac\'{a} Grant 4720-11.

\end{acknowledgments}

\end{document}